\documentclass[a4paper,fleqn]{cas-dc}
\usepackage[authoryear,longnamesfirst]{natbib}
\usepackage{CJKutf8}
\usepackage{natbib}
\setcitestyle{numbers,square}
\def\tsc#1{\csdef{#1}{\textsc{\lowercase{#1}}\xspace}}
\tsc{WGM}
\tsc{QE}
\tsc{EP}
\tsc{PMS}
\tsc{BEC}
\tsc{DE}

\begin{document}
\begin{CJK}{UTF8}{gbsn}
\let\WriteBookmarks\relax
\def\floatpagepagefraction{1}
\def\textpagefraction{.001}



\title [mode = title]{Interface and spectrum multiplexing ptychographic reflection microscopy}                      



%
\author[1]{Yun Gao}[style=chinese]





\affiliation[1]{ organization={School of Physics and Wuhan National Laboratory for Optoelectronics, Huazhong University of Science and Technology},   
    city={Wuhan},
    postcode={430074}, 
    country={China} }

\author[1] {Qijun You}[style=chinese]

%
\author[1,2]{Peixiang Lu}[style=chinese]

\author[1]{Wei Cao}[style=chinese]
\cormark[1]
\ead{weicao@hust.edu.cn}


\affiliation[2]{organization={Optics Valley Laboratory},  
	city={Wuhan},
	postcode={430074}, 
	country={China}}


\cortext[cor1]{Corresponding author}


\begin{abstract}
	Reflective ptychography is a promising lensless imaging technique with a wide field of view, offering significant potential for applications in semiconductor manufacturing and detection. However, many semiconductor materials are coated with different layers during processing, which leads to the reflected diffraction light being a coherent superposition of multiple light beams. Traditional phase recovery methods often overlook the multi-layered nature of these materials, resulting in artificial errors and, in some cases, failures in image reconstruction. This limitation has hindered the broader application and adoption of reflection ptychography. We propose and experimentally demonstrate an innovative interface and spectrum multiplexing ptychographic reflection microscopy here. By employing multi-wavelength light as the illumination source, our approach enables the accurate extraction of the wavelength-dependent surface structure imaging and topography mapping of materials in a single experiment. This advancement offers a reliable technique for element-specific detection of semiconductor materials using tabletop extreme ultraviolet light sources in the future.
 
\end{abstract}




\maketitle

\section{Introduction}
	As one of the lensless imaging techniques, coherent diffraction imaging(CDI) simultaneously recovers both the intensity and phase information of samples\cite{miao1}, making it highly valuable in biomedicine, materials science, and electronic devices research\cite{2, 3, 4, 5, 12}. By combining the CDI technique with advanced light sources that have the potential for high spatial resolution, a certain depth of nondestructive penetration, and chemical sensitivity\cite{16,7,17,20}, it is possible to present new opportunities for advancing nanomaterials and nanotechnology. Ptychography greatly improves the robustness based on the original coherent diffractive imaging through the overlap between different probing points. The increasing redundancy allows fewer prior illumination conditions, and decoupling of different incoherence illumination, object, and noise modes\cite{8,9,10}. 
	
	Reflection ptychography has further expanded the applications of this lensless technique, particularly for relatively thick, opaque samples and high-contrast phase imaging applications. In contrast to transmission ptychography, reflection ptychography provides valuable insights into surface morphology, elemental composition specificity, and buried interface structures\cite{11,13,14,6,7}.  Particularly relevant in the semiconductor industry and nanofabrication technologies,  utilizing ptychography in the extreme ultraviolet（EUV）band enables nanometer-scale spatial resolution,  non-destructive examination with certain penetrating depths of silicon-based chips\cite{31,32,19}. And lensless reconstruction avoids the optical component processing challenges in the EUV band. Therefore, the continued development of reflection ptychography is essential for advancing these technologies.

	However, our investigation has identified a significant challenge in imaging various practical samples, such as silicon-based chips and biomaterials, using ptychographic reflection microscopy. Due to the intrinsic properties of these materials, along with their fabrication processes, the probe does not fully reflect from the top surface. This causes portions of the transmission probe to reflect from each interface and coherently superimpose at the detector interface, complicating the diffraction information obtained and hindering accurate image reconstruction. While, some approaches aiming to utilize multiple spatially or spectrally separated illuminations, each of which still maintains a certain degree of coherence, can reconstruct different object functions and illumination functions as different modes. These methods typically rely on the mutual incoherence between different illumination modes\cite{8,9,10}. In contrast, the waves reflected from different interfaces cannot typically be treated as separate incoherent modes, as is the case in temporally multiplexed multibeam ptychographic imaging \cite{33}. Furthermore, the intermingling of information from various interfaces complicates the reconstruction process. Consequently, current multi-mode ptychography algorithms, which could reconstruct multiple incoherent modes simultaneously, fail to effectively decouple multiplexed interface information within this reflection geometry. These complexities in reflection ptychography frequently fail reconstruction, posing a significant challenge and hindering the widespread adoption of reflection ptychography. Thus, more rigorous data processing and reconstruction processes for reflection ptychography are necessary.

	To solve the above problems, this paper proposes an imaging scheme “interface and spectral multiplexing ptychography reflection microscopy (PISM)”, which firstly takes into account the interference of multiple reflected waves due to the intrinsic characters of the sample or layered structures and the resulting coupling of information, and realizes the precise imaging of the surface structure of the sample and the spectral response analysis. PISM can also be applied to single-shot ptychographic imaging\cite{22,21}with a wide field of view, without needing to separately account for aperture distances to ensure spatial incoherence between beams. We demonstrated the effectiveness of this method using measured data from two respective samples, a negative 1951 resolution test chart, and an abandoned chip.
	
	Our proposed imaging system effectively avoided the artifacts often found in reflection-based imaging, accurately reconstructed the sample, and successfully acquired its spectral and spatial multiplexing information. This advancement overcomes a key barrier to realizing widespread applications of reflectance-based imaging, providing technical support for surface morphology diagnostics, nondestructive testing, and material identification. Our proposed scheme can be extended to EUV sources and has potential applications in realizing full 3D dynamic ptychographic imaging in nanotechnology.

\section{Results}
\subsection{Interfaces and spectrum multiplexing ptychographic reflection microscopy methodology}

	Considering two mutually coherent beams, $\mit\Psi_{1}(r)$ and $\mit\Psi_{2}(r)$,  scatterd from different sample interfaces and propogating to the far field for detection. The diffraction pattern is recorded as:

\begin{equation}
\begin{split}
	I(k)  &= \left|\mathcal{F}[\mit\Psi(r)]\right|^{2}  \\
&=\left|\mathcal{F}[\mit\Psi_{1}(r)+\mit\Psi_{2}(r)]\right|^{2}\\
&=\left|\widetilde{\mit\Psi}_{1}(k)\right|^{2}+\left|\widetilde{\mit\Psi}_{2}(k)\right|^{2}+\widetilde{\mit\Psi}_{1}(k)\widetilde{\mit\Psi}_{2}(k)^{\ast}+\widetilde{\mit\Psi}_{1}(k)^{\ast}\widetilde{\mit\Psi}_{2}(k)
\end{split}
\end{equation} 

where $\mathcal{F}$ is the 2D Fourier transform operator and $\widetilde{\mit\Psi}$ is the form of $\mit\Psi$ after Fourier transform. Mutual interference between the adjacent beams introduces two interference terms in equation 1. According to Wolf’s theory of coherent modes\cite{Wolf:82}, a partially coherent wave can be modeled as a set of orthogonal coherent modes, only incoherent patterns can be reconstructed. Therefore, the interference terms need to be eliminated.

\begin{figure*}[ht]
	\centerline{\includegraphics[width=0.9\linewidth]{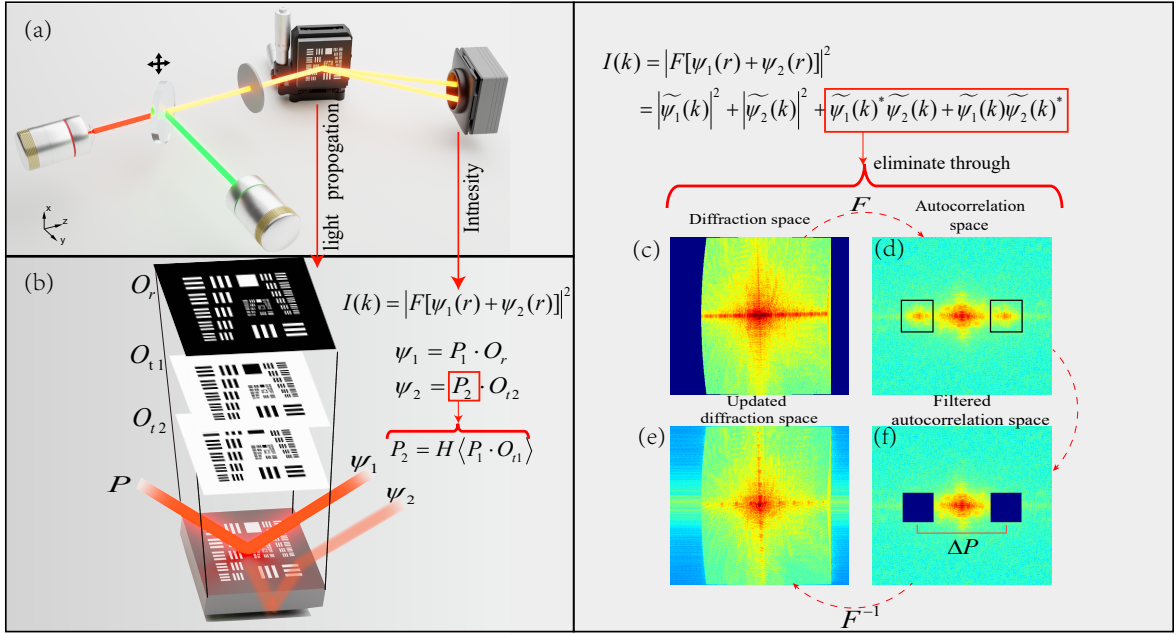}}
	\caption{\label{fig1}\footnotesize  Geography and methodology illustration of the reflective ptychography microscopy. (a) Geography of the reflective ptychography microscopy. (b) Light propogation in the sample. After the tilt correction\cite {23}, diffractive intensity is the squares of the Fourier transform of the sum of reflected waves $\Psi_{1}$ and  $\Psi_{2}$. Each of the reflected waves is expressed based on the probe's function $P$, the sample's reflection function $O_{r}$, and transmission functions $O_{t1}$, $O_{t2}$. \textit{F} is  the 2D Fourier transform operator and $\widetilde{\mit\Psi}$ is the form of $\mit\Psi$ after Fourier transform. $ H<>$ is the tilt angular spectrum propagator.   (c)-(f) A flow chart of the autocorrelation filtering process: (b) The diffraction data after the tilt correction. (c) Data after a Fourier transform from the data in (b). (d) Data after removing the cross-correlation terms in (b). (e) Data after an inverse Fourier transforming the data in (d).}
\end{figure*}

	For a sample with a thickness of the substrate, beams reflected from different interfaces will have a certain spatial translation with each other, just shown as [Fig.~\ref{fig1}.b]). Applying this space interval, we eliminate the additional two terms in equation 1 by auto-correlation filtering\cite {25}. The whole data processing is shown in [Fig.~\ref{fig1}.c-f]. Operating the Fourier transform to the diffraction intensity, the cross-correlation terms ( framed in black in [Fig.~\ref{fig1}.d]), originating from the interference terms are separated from the auto-correlation terms. Based on this, we can remove the isolated cross-correlation terms and then do the inverse Fourier transform, resulting in diffraction intensity free from the interference terms. In this way, the processed diffraction pattern can be regarded as the incoherent superposition of multiple beams reflected from various interfaces, and different modes can be decoupled.

	The propagation process of light in the sample is shown in [Fig.~\ref{fig1}.b]. Note in the reflected wave $\Psi_{2}$, the probe passes the sample twice complicating the reconstruction of probe's and sample's functions. We introduce the reflection ptychographic interface and spectrum multiplexing iteration method (PISM), which effectively decouples the information from different interfaces combining with the off-axis light propagation and principle of multi-slice approach\cite{29}. In [Fig.~\ref{fig1}.b],  $ \mit\Psi_{1} $ is the beam that is reflected directly from the top and it can be expressed as: $P\cdot O_{r}$, where $ O_{r} $ is uppermost reflection function, $ P $ is the incident light. A portion of the light that does not reflect passes through the top surface of the sample, reflects back from the substrate and passes through the sample's top surface for another time. This is the second beam $ \mit\Psi_{2} $ and its expression can be written as: $ \mit\Psi_{2} =  H\left\langle P \cdot O_{t1}\right\rangle  \cdot O_{t2}$, where $ H $ represents  the tilt angular spectrum propagation in the substrate and $ O_{t1}$ the first transmission function, $ O_{t2}$the second transmission function. Finally, these two beams interfere with each other and scatter onto the detector's receiving surface. Now, we rewrite the diffraction pattern as
\begin{equation}
\begin{split}
	I(k) &= \left|\mathcal{F}[\mit\Psi(r)]\right|^{2} \\
& =  \left|\mathcal{F}[\mit\Psi_{1}(r)+\mathcal{F}[\mit\Psi_{2}(r)]\right|^{2}\\
& = \left|\mathcal{F}(P\cdot O_{r})+\mathcal{F}(H\left\langle P \cdot O_{t1}\right\rangle \cdot  O_{t2})\right|^{2}\\
& = \left|\mathcal{F}(P\cdot O_{r})\right|^{2}+\left|\mathcal{F}(H\left\langle P \cdot O_{t1}\right\rangle \cdot  O_{t2})\right|^{2}
\end{split}
\end{equation}  

	In the multi-slice approach, the transmitted probe propagated for a distance can be seen as a new probe in the form of  $ H\left\langle P \cdot O_{t1}\right\rangle $ for the next transmission. In that case, in each iteration, $ O_{r},O_{t2}$ are recovered directly from $\Psi_{1}$,$\Psi_{2}$ ​respectively. $O_{t1}$can be attained through the other nested update based on the principles of the multi-slice approach. At the end, $P, O_{r}, O_{t1}, O_{t2}$ are recovered simultaneously after certain number of iterations. 
	As for the multi-wavelength diffraction intensity $ I_{m} $ is the integral over different wavelengths:
\begin{equation}
\begin{split}
	I(k)_{m}=\sum_{i=1}^{m} \left|\mathcal{F}(P^{m}O^{m}_{r})\right|^{2}+\left|\mathcal{F}(H^{m}\left\langle P ^{m}\cdot O^{m}_{t1}\right\rangle \cdot  O^{m}_{t2})\right|^{2}
\end{split}
\end{equation}  

\subsection{Interfaces and spectrum multiplexing ptychographic reflection microscopy setup}
	The schematic of our interfaces and spectrum multiplexing ptychographic reflection microscopy is shown in [Fig.~\ref{fig1}.a]. The collimated incident light passes through a 200nm precise pinhole and illuminates the sample which is mounted on a two-dimensional translation stage. The precise pinhole provided simple geometrical constraints as real space’s support. It must be carefully positioned to avoid obstructing the light path in the reflection ptychography setup. Scattered light from objects, which is the coherent superposition of multiple beams reflecting from different interfaces, was measured by a CCD detector placed 0.08 m away from the sample. To demonstrate the developed approach mentioned above, firstly, we performed experiments with a 633nm monochromatic light (continuous-wave red laser) on a negative USAF1951 resolution test chart and an abandoned chip. We are particularly interested in the sample's specific response to the spectrum, a phenomenon commonly seen in chip inspection and the imaging of biomaterials\cite{15}. To investigate this, we combine 633 nm monochromatic light with another 532 nm monochromatic light and illuminate the sample simultaneously, as shown in [Fig.~\ref{fig1}.a]. The red and green beams are combined using a beam splitter and then directed onto the sample through a precise pinhole. In all experiments in this paper, the incident angle of the light beam on the sample is 45 degrees. The recorded diffraction patterns were captured by a detector with 2048×2048 pixels and spliced from three diffraction patterns, each acquired with different exposure times,  to enlarge the dynamic range and increase the patterns’ signal-to-noise ratio. The scanning form is an S-type detour. At each moving step, the fixed step size is superimposed with a random offset of about 10 percent of the aperture size to avoid the occurrence of periodic artifacts in the reconstructed image and enhance the reconstruction speed. In the monochromatic interfaces multiplexing reflection ptychography experiments, the overlap rate between adjacent scan locations is 70-80 percent. To meet the redundancy requirement for the recovery of multiple modes, an overlap rate of 85 percent was selected for the additional spectrum multiplexing experiments. The former collected 121 images and the latter 225 images.

\subsection{Negative 1951 resolution test chart}
\begin{figure*}[ht]
	\centerline{\includegraphics[width=1\linewidth]{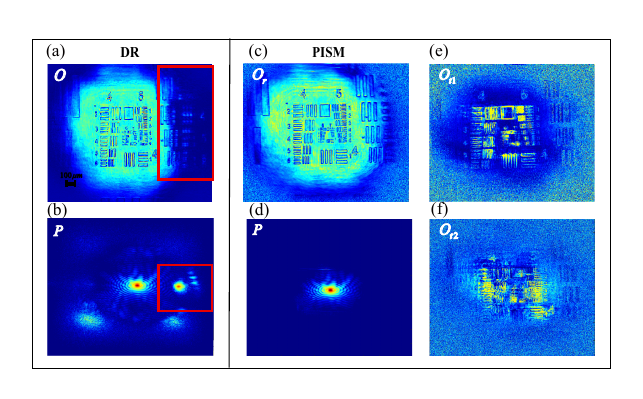}}
	\caption{\label{fig2}\footnotesize Reconstructions of negative reflective 1951 resolution test chart for the 633nm monochromatic source. (a)-(b) Reconstructions by sDR. (c)-(f) Reconstructions by PISM: (c)the recovered amplitude of the sample reflection function from$\left| \mit\Psi_{1}\right| $, (d) the recovered probe, (e)-(f)the recovered amplitude of the sample transmission function from $\left| \mit\Psi_{2}\right| $   }
\end{figure*}

	We collected diffraction data from a negative 1951 resolution test chart using a reflection-based coherent diffraction imaging geometry with a 633nm monochromatic light. The reconstructed amplitudes of the imaged chart with the usual ptychography method—Douglas-Rachford algorithm(sDR)\cite{28} and PISM algorithm are respectively shown in the [Fig.\ref{fig2}.a,c].
	As illustrated in [Fig.\ref{fig2}.a], after correcting the tilt diffraction data\cite {23}, the recovered object image by sDR and the diffraction data after Fourier transform shown in [Fig.\ref{fig1}.d] exhibit strong interference between two beams originating from two different interfaces, and the corresponding two probes are simultaneously reconstructed, as shown in [Fig.\ref{fig2}.b]. These erroneous features are highlighted by red lines in [Fig.~\ref{fig2}.a,b]. It is evident that the recovered object consists of a mixture of the transmission function (right part of [Fig.\ref{fig2}.a]) and the reflection function (left part of [Fig.\ref{fig2}.a]). After the improved reconstruction procedures, the error is corrected and all the information $P$, $O_{r}$, $O_{t1}$, $O_{t2}$ are differentiated and recovered as shown in the [Fig.~\ref{fig2}.c-f]. Since the sample is a negative test chart, meaning that a higher proportion of light intensity is reflected back while less is transmitted into the substrate, the transmitted light intensity gradually decreases during propagation. Consequently, the signal-to-noise ratio of the scattered signals containing transmission functions decreases, leading to poorer performance in the two reconstructed transmission images([Fig.\ref{fig2}.e] and [Fig.\ref{fig2}.f]) compared to the reflection image([Fig.\ref{fig2}.c]), particularly at the edges of the recovered images. This degradation occurs because, after the probe with an Airy spot shape has propagated over a certain distance, the intensity at the edges is significantly attenuated. However, the reconstructed transmission result at the first interface, $ P_{t1}$ can still guarantee a certain resolution. Under the conventional overlap ratio typically used in traditional ptychographic imaging, our method can recover richer and more accurate sample information, highlighting its superior robustness.
	
\begin{figure}[ht]
	\centerline{\includegraphics[width=1\linewidth]{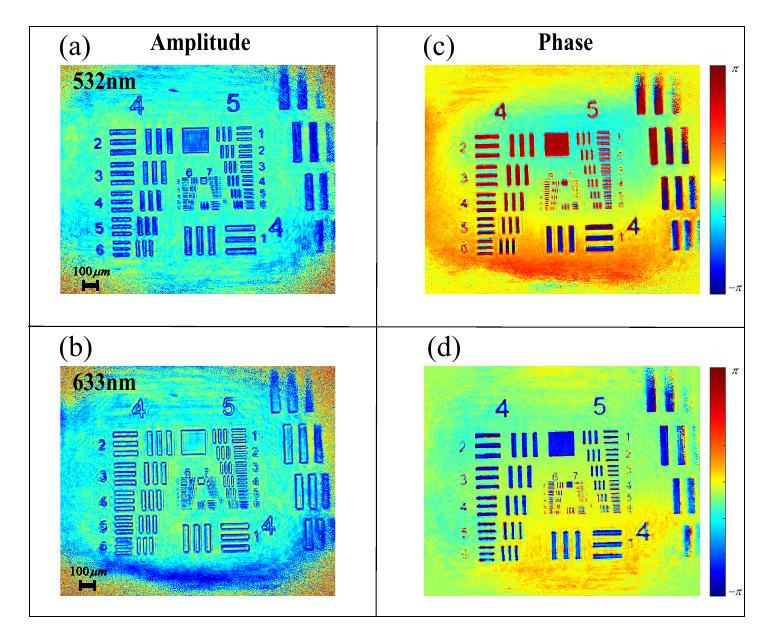}}
	\caption{\label{fig3}\footnotesize Reconstructions of reflective 1951 resolution test chart for multi-wavelength source by PISM algorithm. (a) The recovered amplitude of object reflection function for 532nm light. (b) The recovered amplitude of object reflection function for 633nm light.  (c) The recovered phase of object reflection function for 532nm light. (d) The recovered phase of object reflection function for 633nm light. }
\end{figure}

	   We extended the technology to include spectral multiplexing functionality using an additional 633 nm laser. The data processing in [Fig.\ref{fig1}.c-f] and the new image reconstruction procedure PISM are operated in the dual-wavelength experiment, in which the 532nm and 633nm narrow continuous wavelength lasers are combined to illuminate the chart at the same time. The solution to different wavelengths is consistent with the $ \mit\Psi_{1}, \mit\Psi_{2} $ which can be considered two incoherent modes, so the diffraction intensity can be regarded as the incoherent superposition of four modes in total. The recovered reflection amplitude and phase images for dual laser sources are given in [Fig.~\ref{fig3}]. It can be seen from three groups of test chart reconstructions that the experimental imaging effect of multi-wavelength is nearly the same as that of monochromatic ones, as shown in [Fig.~\ref{fig2}]. The resolution of the sample is determined by both numerical aperture(NA) and wavelength. Due to the shorter wavelength, the resolution of green light is higher than that of red light, which aligns with the recovered results shown in [Fig.~\ref{fig3}.], that the 6,7 groups of test chart in [Fig.~\ref{fig3}.b] are slightly fuzzier than them in [Fig.~\ref{fig3}.a]. The recovered images of the test chart with red light exhibited slightly lower contrasts between the coating and the substrate than the recovered images with green light but aligned with the results of the red monochromatic experiment. Under the illumination of 45° incidence of red light, the reflectivity of chromium coating is lower than that of green light, so the contrast is relatively poor. 

\subsection{Abandoned chip}
\begin{figure*}[ht]
	\centerline{\includegraphics[width=1\linewidth]{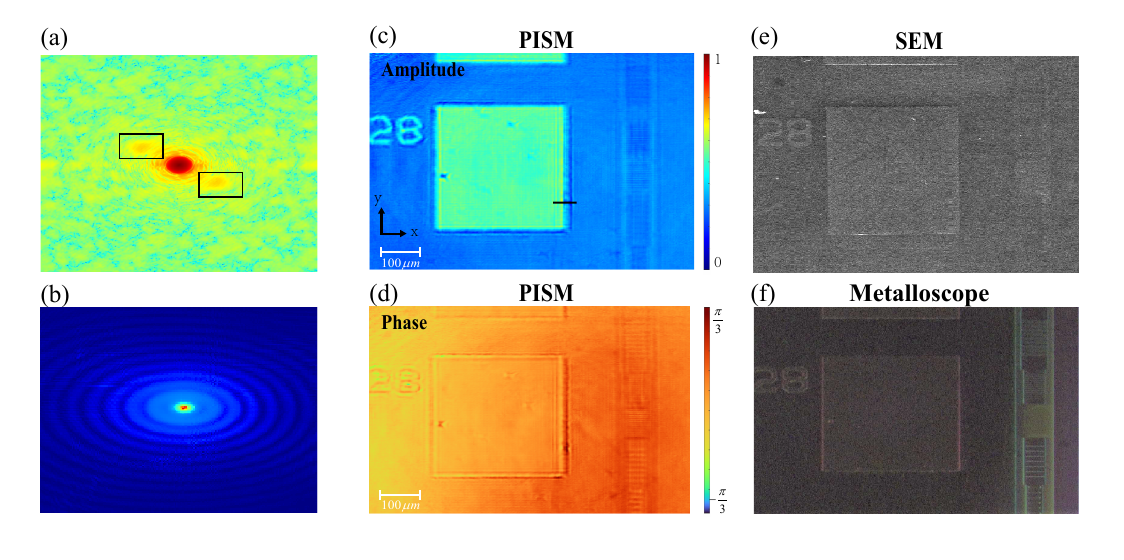}}
	\caption{\label{fig4}\footnotesize Data and reconstructions of the abandoned chip for 633nm monochromatic source with  PISM algorithm. (a) Data after the Fast Fourier transform of the intensity. (b) The recovered probe by the PISM algorithm. (c) The recovered amplitude of the object reflection function by the PISM algorithm. (d) The recovered phase of object reflection function by PISM algorithm. (e)Scanning electron microscopy (SEM) image of the same area on the chip. (f)Metallographic of the same area on the chip. }
\end{figure*}

The same reflection ptychography imaging geometry is applied to an abandoned chip sample with some etched patterns on the surface. Both in the monochromatic experiment and the dual-wavelength experiment, after operating the Fourier transform to the scatted intensity, in the auto-correlation space, there are two additional cross-items transformed from the interference items, which have been marked in black boxes in  [Fig.~\ref{fig4}.a], this means the similar interference probably between the surface reflection beam and the substrate reflection beam exists.  Therefore, data was processed identically to the last experiment and then, we applied the developed algorithm to reconstruct the amplitude and phase information of the chip and probe. The recovered probe's amplitude and chip surface structure with a 633nm light source are shown in  [Fig.~\ref{fig4}.b-d]. 

\begin{figure}[ht]
	\centerline{\includegraphics[width=1\linewidth]{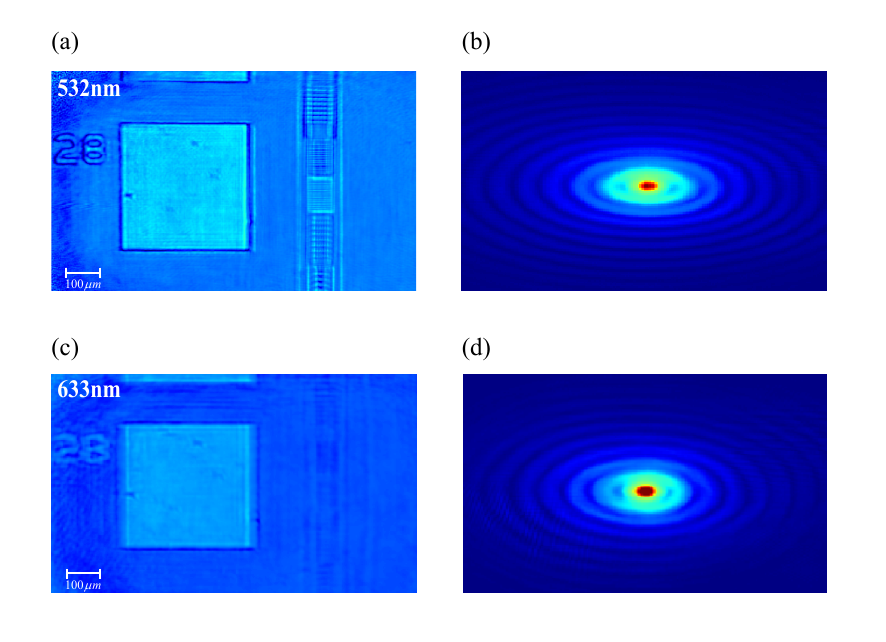}}
	\caption{\label{fig5}\footnotesize Reconstructions of abandoned chip for multi-wavelength source by PISM algorithm. (a)(b) The recovered amplitude of object reflection and probe function with a 532nm light source. (c)(d) The recovered amplitude of object reflection and probe function with a 633nm light source.}
\end{figure}

The reconstructed amplitudes of the probes and objects in the dual-wavelength experiment are shown in [Fig.~\ref{fig5}]. Notably, the raster-like structure etched on the right side of the chip reflected the weaker red light signal at an incident angle of 45°,  resulting in the reconstruction of red light illumination being less pronounced compared to the reconstruction of green light illumination. These results demonstrate that the chip exhibits different reflected intensity distributions depending on the wavelength used. Furthermore, in this experiment, the object function reconstructed in the red light mode shows a relatively lower resolution compared to the result obtained in the aforementioned red light monochromatic experiment. This can be attributed to the fact that more modes share the signal-to-noise ratio in the dual-wavelength experiment. Additionally, under 45 degrees tilt illumination, the chip's spectral response to green light is significantly stronger than its response to red light, which is evident in the diffraction data, resulting in a higher signal-to-noise ratio for the green light mode compared to the red light mode.


\section{Discussion}
 	
    
      We compare our reconstructed images to the metallographic microscope and Scanning electron microscopy(ZEISS EVO 10) at the same area of the chip. All results are in general agreement except for some raster-like structures on the right side of the field of view. These structures are discernable in reconstructed amplitude and phase of ptychography imaging but cannot be distinguished very well in SEM due to the concession to the field of view. 

	 Previous work using the autocorrelation filtering process has shown that the process of diffraction data has no noticeable effect and information loss on the reconstructed resolution of the object function\cite {25}. To numerically characterize the resolution of the microscope, we chose a profile in x-axial in [Fig.~\ref{fig4}.c] and measured the 10\%-90\% width of it. Specific analysis is added to the APPENDIX A.

\begin{figure}[ht]
	\centerline{\includegraphics[width=1\linewidth]{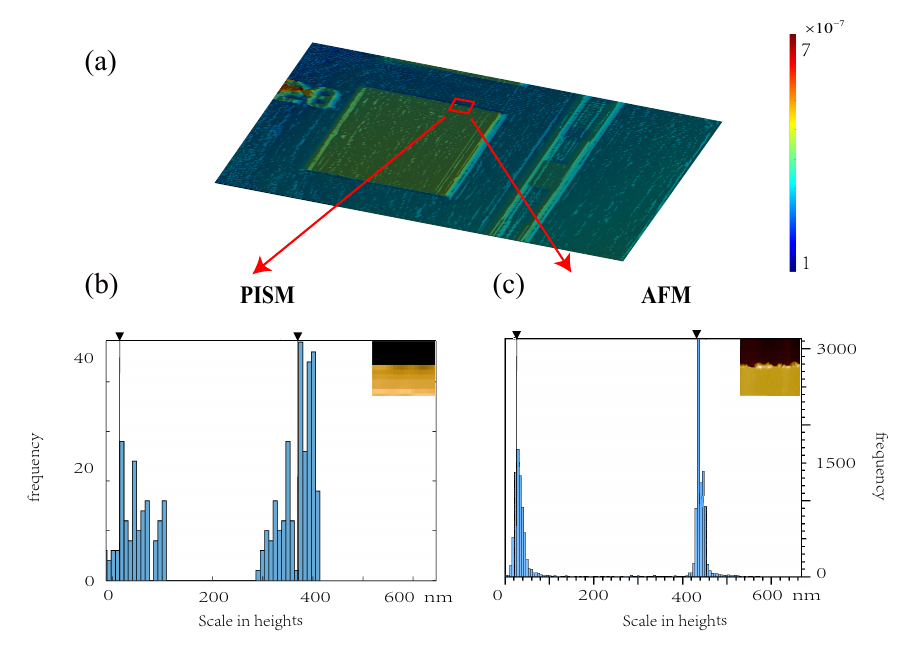}}
	\caption{\label{fig6}\footnotesize Height analysis based on interfaces and spectrum multiplexing ptychographic reflection microscopy and AFM. (a) Topography mapping by the interfaces and spectrum multiplexing ptychographic reflection microscopy.  (b)Histogram of the interfaces and spectrum multiplexing ptychographic reflection microscopy. (The small picture in the upper right corner is the reconstruction in [Fig.~\ref{fig5}] (c)Histogram of AFM (small picture in the upper right corner is the AFM image )  }
\end{figure}

     Reflectance ptychography is a highly phase-sensitive technique enabling quantitative surface profiling. As an extension of reflectance ptychography, PISM effectively preserves these properties. To assess the axial resolution of PISM, prior information such as wavelength, angle of incidence, and sample composition is utilized to extract height information from the reconstructed phase. The wavelength-dependent topography mapping of the chip made from the reconstructed phase of 532nm mode in the dual-wavelength experiment is shown in  [Fig.~\ref{fig6}.a]. Due to the lack of absolute phase reference, relative height information can be obtained. The specific heights of the surface structures are still accessible from it. We extract the height information of the feature area that has a stepped height transition (marked with the red box in [Fig.~\ref{fig6}.a] using an atomic force microscope(AFM)(Nanosurf Flex-Mount), and compared it with the reconstructed height distribution using reflection ptychography. The histograms for the height values within this feature area are shown in [Fig.~\ref{fig6}.b] and [Fig.~\ref{fig6}.c]. Both histograms have a good agreement, indicating that the height difference of this stepped structure is between 360nm - 400nm.

	In sum, we have presented a novel interface and spectrum multiplexing ptychographic reflection microscopy technique that, for the first time, simultaneously accounts for complex interferences from various sample interfaces and the spectral characteristics of the sample. Two different samples were selected for separate imaging experiments, demonstrating that PISM can reliably and accurately reconstruct the reflective intensity of a sample and precisely characterize its surface topography. It is believed that this scheme can expand the application areas of reflection lensless phase imaging, and foreseeably, its combination with extreme ultraviolet light sources will play an important role in the exploration of high spatial and temporal resolution.

\section{Material and Methods}
\subsection{ Auto-correlation filtering }

	 In autocorrelation space, that is, after the Fourier transform of the diffraction intensity, the information is written as:
\begin{equation}
\begin{split} 
\mathcal{F}[I(k)]&={{\mit\Psi}_{1}(r)} \star {{\mit\Psi}_{1}(r)}+{{\mit\Psi}_{2}(r)} \star {{\mit\Psi}_{2}(r)}+\\
  &{{\mit\Psi}_{1}(r)} \star  {{\mit\Psi}_{2}(r)^{\ast}}+{{\mit\Psi}_{1}(r)^{\ast}} \star {{\mit\Psi}_{2}(r)}
\end{split}
\end{equation} 
where $\star$ is the 2D auto-correlation operator and ${}^{\ast}$ indicates the complex conjugatation.
	
	Express $ \mit\Psi_{1}(r)$ as $ \mit\Psi_{1}(r)={P_{1}(r-R_j)O_{1}(r)}$ and $ \mit\Psi_{2}(r)$ as $ \mit\Psi_{2}(r)={P_{2}(r-R_j-\Delta r)O_{2}(r)}$, where $ \mit\Psi_{1}(r)$, $\mit\Psi_{2}(r)$ are two exited  waves. $O_{1}(r)$,  $O_{2}(r)$ are object functions in scan position$R_j$. $P_{1}$, $P_{2}$ are probe functions of $\mit\Psi_{1}(r)$ and $\mit\Psi_{2}(r)$. $\Delta r$ is the distance between two exit waves.

	 Then, equation 4 can be further written as:
\begin{small}
\begin{equation}
\begin{split} 
\mathcal{F}[I(k)]&={P_{1}(r-R_j)O_{1}(r)} \star {P_{1}(r-R_j)O_{1}(r)}+\\&{P_{2}(r-R_j-\Delta r)O_{2}(r)} \star {P_{2}(r-R_j-\Delta r)O_{2}(r)}+\\
  &{{{P_{1}(r-R_j)O_{1}(r)} \star  {P_{2}(r-R_j-\Delta r)O_{2}(r)}}^{\ast}}\otimes{\delta(r+\Delta r)}+
\\&{{P_{1}(r-R_j)O_{1}(r)} \star {P_{2}(r-R_j-\Delta r)O_{2}(r)}}\otimes{\delta(r-\Delta r)}
\end{split}
\end{equation} 
\end{small}
where $\otimes$ is the 2D convolution operator.	
	
	As shown in [Fig.~\ref{fig1}.d], due to the spatial separation $ \Delta x$ in the X-axis direction between $\Psi_{1}$ and $\Psi_{2}$ in real space, there is a corresponding separation  $ \Delta P $ between the center of two cross-correlation terms and auto-correlation terms in auto-correlation space\cite{25,26,27}. The mapping of $ \Delta x $ over the number of pixels can be expressed as $ \Delta P = \Delta xD/\lambda Z$, where Z is the distance between the object and the camera, and D is the camera size. By utilizing the numerical relationship, the spatial distribution of the exit waves can be determined.   Removing the isolated area shown in the auto-correlation space in the [Fig.~\ref{fig1}.d](two blackened squares) and operating the inverse Fourier transform for another time, we can filter out the excess interference terms. [Fig.~\ref{fig1}.c-f] are the whole process of data processing of the autocorrelation filter after the tilt correction. Through autocorrelation filtering, corrected diffraction patterns can be transformed into the sum of incoherent terms, and then the magnitude and phase information of the probe and object can be recovered in different modes simultaneously.

\begin{figure*}[ht]
	\centerline{\includegraphics[width=1\linewidth]{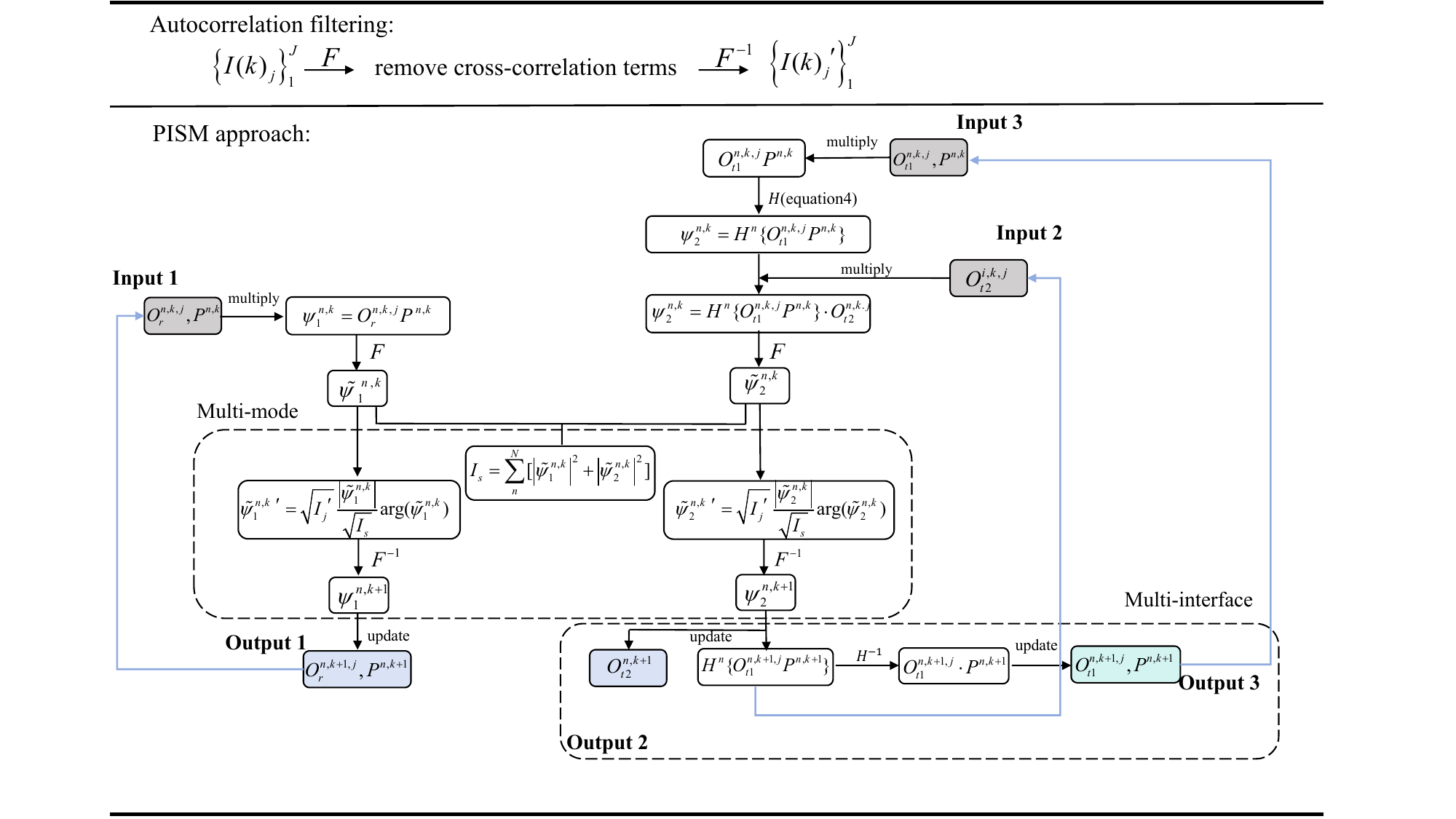}}
	\caption{\label{fig7}\footnotesize A flow chart of the algorithm used by the interfaces and spectrum multiplexing ptychographic reflection microscopy.n denotes the ordinal number of wavelength modes. k denotes the ordinal number of interactions. j denotes the ordinal number of the scan points.}
\end{figure*}

\subsection{ Off-axis light propagation }

	In the multi-slice approach, the propagated exit wave of the first transmission $ H\left\langle P \cdot O_{t1}\right\rangle $ is computed as the incident probe for the next transmission. In terms of the off-axis light propagation between parallel layers of the sample with an arbitrary angle, a numerical method named Translated-angular spectrum method(Translated-ASM)\cite{30} was used to accurately calculate diffraction fields in the disparate interfaces. The propagation of the exit wave function $ P\cdot O_{t1}$ can be described by the Translated-angular spectrum function as:
 
\begin{equation}
\begin{split}
	\mit\Psi_{0}(x^{'},y^{'};Z)=\mathcal{F}^{-1}(\mathcal{F}(E_{i}(x,y;0))exp(i \widetilde{k}_{z}R)[k_{x},k_{y}])
\end{split}
\end{equation} 
where x,y,z are original coordinates, $x^{'},y^{'},z^{'}$ are shifted new coordinates, $k_{x},k_{y}$ are the spatial frequency variables and $\widetilde{k}_{z}R $ is the projection of an angular spectrum component onto the tilted axis.
In each iteration, $ O_{r},O_{t2}$ ,shown in [Fig.~\ref{fig1}.a] are recovered directly from $\Psi_{1}$,$\Psi_{2}$ ​respectively. Since $ O_{t1}$ is related to the probe of $\mit\Psi_{2}$, $O_{t1}$ can be attained through the other nested update based on the principles of the multi-slice approach. At the end, $P,O_{r},O_{t1},O_{t2}$ are recovered simultaneously after certain number of iterations.  

\subsection{ Iterative recovery process }

	As for the iterative recovery process, because the reflection-mode ptychography geometry uses the precise pinhole as support, the probe would be a non-standard circular diffraction, so we choose the Douglas-Rachford algorithm(sDR), which has better performance in practical applications, especially for relatively complex situations with limited prior information of the probe. All the phase iterative recovery procedure is detailed in[Fig.~\ref{fig7}].

\subsection{Materials }
\begin{figure}[ht]
	\centerline{\includegraphics[width=1\linewidth]{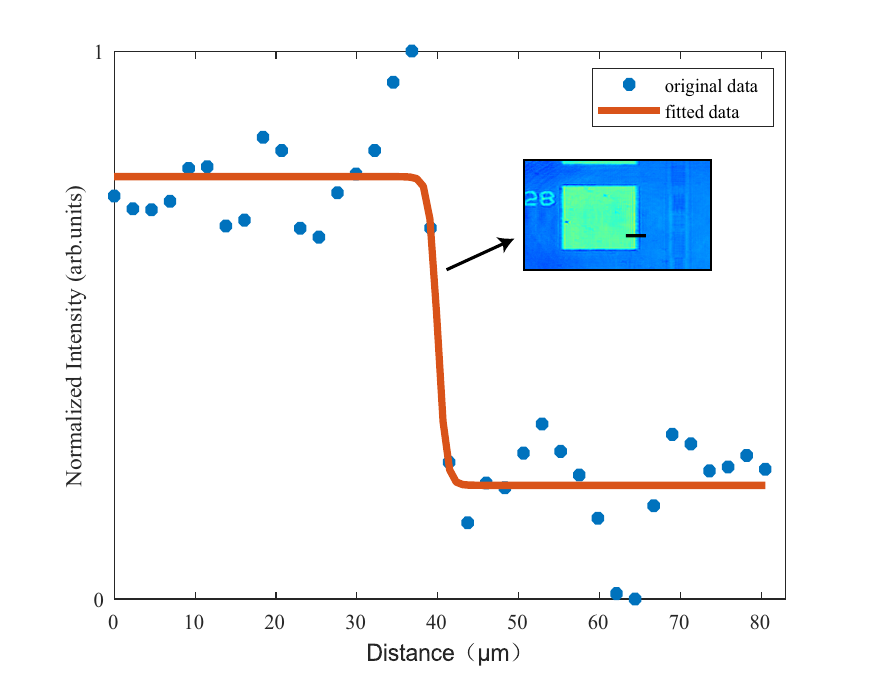}}
	\caption{\label{fig8}\footnotesize Resolution evaluation of the interfaces and spectrum multiplexing reflection microscopy. }
\end{figure}

	The abandoned chip sample is a silicon-based chip,  which is a bad sheet produced during the chip production process, and the surface has some lithographic nanometer patterns.

\*{APPENDIX A：}
	To numerically evaluate the resolution of the Interface and spectrum multiplexing Ptychographic reflection microscopy, we choose a rising edge (shown in the inset of [Fig.~\ref{fig8}]) in the central of the reconstructed amplitude of the abandoned chip to avoid the poor reconstruction quality due to the inadequate robustness near the edges. The featured profile has a width of 3.9 nm.

\subsection*{Funding}
The National Natural Science Foundation of China (Grant Nos. 12274158 and 12021004), and the Open Foundation Project of Hubei Key Laboratory of Optical Information and Pattern
Recognition of Wuhan Institute of Technology (Grant No. 202304).

\bibliographystyle{unsrt}

\bibliography{manu.bib}

\begin{thebibliography}{10}

\bibitem{miao1}
Jianwei Miao, Pambos Charalambous, Janos Kirz, and David 
\newblock Extending the methodology of x-ray crystallography to allow imaging
  of micrometre-sized non-crystalline specimens.
\newblock 400(6742):342--344, 1999.

\bibitem{2}
David Shapiro, Pierre Thibault, Tobias Beetz, Veit Elser, Malcolm Howells,
  Chris Jacobsen, Janos Kirz, Enju Lima, Huijie Miao, Aaron~M. Neiman, and
  David Sayre.
\newblock Biological imaging by soft x-ray diffraction microscopy.
\newblock 102(43):15343--15346, 2005.

\bibitem{3}
Changyong Song, Raymond Bergstrom, Damien Ramunno-Johnson, Huaidong Jiang,
  David Paterson, Martin~D De~Jonge, Ian McNulty, Jooyoung Lee, Kang~L Wang,
  and Jianwei 
\newblock Nanoscale imaging of buried structures with elemental specificity
  using resonant x-ray diffraction microscopy.
\newblock 100(2):025504, 2008.

\bibitem{4}
Ashish Tripathi, Jyoti Mohanty, Sebastian~H. Dietze, Oleg~G. Shpyrko, Erik
  Shipton, Eric~E. Fullerton, Sang~Soo Kim, and Ian McNulty.
\newblock Dichroic coherent diffractive imaging.
\newblock 108(33):13393--13398, 2011.

\bibitem{5}
Ofer Kfir, Sergey Zayko, Christina Nolte, Murat Sivis, Marcel Möller, Birgit
  Hebler, Sri Sai Phani~Kanth Arekapudi, Daniel Steil, Sascha Schäfer, and
  Manfred 
\newblock Nanoscale magnetic imaging using circularly polarized high-harmonic
  radiation.
\newblock 3(12):eaao4641, 2017.

\bibitem{12}
Joanne Marrison, Lotta Räty, Poppy Marriott, and Peter O'Toole.
\newblock Ptychography – a label free, high-contrast imaging technique for
  live cells using quantitative phase information.
\newblock {\em Scientific Reports}, 3(1):2369, 2013.

\bibitem{16}
Dennis~F. Gardner, Michael Tanksalvala, Elisabeth~R. Shanblatt, Xiaoshi Zhang,
  Benjamin~R. Galloway, Christina~L. Porter, Robert Karl~Jr, Charles Bevis,
  Daniel~E. Adams, Henry~C. Kapteyn, Margaret~M. Murnane, and Giulia~F.
  Mancini.
\newblock Subwavelength coherent imaging of periodic samples using a 13.5 nm
  tabletop high-harmonic light source.
\newblock {\em Nature Photonics}, 11(4):259--263, 2017.

\bibitem{7}
Elisabeth~R. Shanblatt, Christina~L. Porter, Dennis~F. Gardner, Giulia~F.
  Mancini, Robert M.~Jr. Karl, Michael~D. Tanksalvala, Charles~S. Bevis,
  Victor~H. Vartanian, Henry~C. Kapteyn, Daniel~E. Adams, and Margaret~M.
  Murnane.
\newblock Quantitative chemically specific coherent diffractive imaging of
  reactions at buried interfaces with few nanometer precision.
\newblock {\em Nano Letters}, 16(9):5444--5450, 2016.

\bibitem{17}
Wilhelm Eschen, Lars Loetgering, Vittoria Schuster, Robert Klas, Alexander
  Kirsche, Lutz Berthold, Michael Steinert, Thomas Pertsch, Herbert Gross,
  Michael Krause, Jens Limpert, and Jan Rothhardt.
\newblock Material-specific high-resolution table-top extreme ultraviolet
  microscopy.
\newblock {\em Light: Science and Applications}, 11(1):117, 2022.

\bibitem{20}
Chang Liu, Wilhelm Eschen, Lars Loetgering, Daniel~S Penagos~Molina, Robert
  Klas, Alexander Iliou, Michael Steinert, Sebastian Herkersdorf, Alexander
  Kirsche, and Thomas 
\newblock Visualizing the ultra-structure of microorganisms using table-top
  extreme ultraviolet imaging.
\newblock 4(1):6, 2023.

\bibitem{8}
Darren~J. Batey, Daniel Claus, and John~M. Rodenburg.
\newblock Information multiplexing in ptychography.
\newblock {\em Ultramicroscopy}, 138:13--21, 2014.

\bibitem{9}
Pierre Thibault and Andreas Menzel.
\newblock Reconstructing state mixtures from diffraction measurements.
\newblock {\em Nature}, 494(7435):68--71, 2013.

\bibitem{10}
Bosheng Zhang, Dennis~F. Gardner, Matthew~H. Seaberg, Elisabeth~R. Shanblatt,
  Christina~L. Porter, Robert Karl, Christopher~A. Mancuso, Henry~C. Kapteyn,
  Margaret~M. Murnane, and Daniel~E. Adams.
\newblock Ptychographic hyperspectral spectromicroscopy with an extreme
  ultraviolet high harmonic comb.
\newblock {\em Optics Express}, 24(16):18745--18754, 2016.

\bibitem{11}
Tao Sun, Zhang Jiang, Joseph Strzalka, Leonidas Ocola, and Jin Wang.
\newblock Three-dimensional coherent x-ray surface scattering imaging near
  total external reflection.
\newblock {\em Nature Photonics}, 6(9):586--590, 2012.

\bibitem{13}
Matthew~D. Seaberg, Bosheng Zhang, Dennis~F. Gardner, Elisabeth~R. Shanblatt,
  Margaret~M. Murnane, Henry~C. Kapteyn, and Daniel~E. Adams.
\newblock Tabletop nanometer extreme ultraviolet imaging in an extended
  reflection mode using coherent fresnel ptychography.
\newblock {\em Optica}, 1(1):39--44, 2014.

\bibitem{14}
Robert~M. Karl, Giulia~F. Mancini, Joshua~L. Knobloch, Travis~D. Frazer,
  Jorge~N. Hernandez-Charpak, Begoña Abad, Dennis~F. Gardner, Elisabeth~R.
  Shanblatt, Michael Tanksalvala, Christina~L. Porter, Charles~S. Bevis,
  Daniel~E. Adams, Henry~C. Kapteyn, and Margaret~M. Murnane.
\newblock Full-field imaging of thermal and acoustic dynamics in an individual
  nanostructure using tabletop high harmonic beams.
\newblock 4(10):eaau4295, 2018.

\bibitem{6}
Michael Tanksalvala, Christina~L. Porter, Yuka Esashi, Bin Wang, Nicholas~W.
  Jenkins, Zhe Zhang, Galen~P. Miley, Joshua~L. Knobloch, Brendan McBennett,
  Naoto Horiguchi, Sadegh Yazdi, Jihan Zhou, Matthew~N. Jacobs, Charles~S.
  Bevis, Robert~M. Karl, Peter Johnsen, David Ren, Laura Waller, Daniel~E.
  Adams, Seth~L. Cousin, Chen-Ting Liao, Jianwei Miao, Michael Gerrity,
  Henry~C. Kapteyn, and Margaret~M. Murnane.
\newblock Nondestructive, high-resolution, chemically specific 3d nanostructure
  characterization using phase-sensitive euv imaging reflectometry.
\newblock 7(5):eabd9667, 2021.

\bibitem{31}
Yifeng Shao, Sven Weerdenburg, Jacob Seifert, H~Paul Urbach, Allard~P Mosk, Wim
\newblock Wavelength-multiplexed multi-mode euv reflection ptychography based
  on automatic differentiation.
\newblock 13(1):196, 2024.

\bibitem{32}
Christina~L. Porter, Michael Tanksalvala, Michael Gerrity, Galen Miley, Xiaoshi
  Zhang, Charles Bevis, Elisabeth Shanblatt, Robert Karl, Margaret~M. Murnane,
  Daniel~E. Adams, and Henry~C. Kapteyn.
\newblock General-purpose, wide field-of-view reflection imaging with a
  tabletop 13\&\#x2009;\&\#x2009;nm light source.
\newblock {\em Optica}, 4(12):1552--1557, Dec 2017.

\bibitem{19}
Stefan Witte, Vasco~T. Tenner, Daniel W.~E. Noom, and Kjeld S.~E. Eikema.
\newblock Lensless diffractive imaging with ultra-broadband table-top sources:
  from infrared to extreme-ultraviolet wavelengths.
\newblock {\em Light: Science and Applications}, 3(3):e163--e163, 2014.

\bibitem{33}
Nathan~J. Brooks, Bin Wang, Iona Binnie, Michael Tanksalvala, Yuka Esashi,
  Joshua~L. Knobloch, Quynh L.~D. Nguyen, Brendan McBennett, Nicholas~W.
  Jenkins, Guan Gui, Zhe Zhang, Henry~C. Kapteyn, Margaret~M. Murnane, and
  Charles~S. Bevis.
\newblock Temporal and spectral multiplexing for euv multibeam ptychography
  with a high harmonic light source.
\newblock {\em Opt. Express}, 30(17):30331--30346, Aug 2022.

\bibitem{22}
Pavel Sidorenko and Oren Cohen.
\newblock Single-shot ptychography.
\newblock {\em Optica}, 3(1):9--14, 2016.

\bibitem{21}
Hannah~C. Strauch, Fengling Zhang, Stefan Mathias, Thorsten Hohage, Stefan
  Witte, and G.~S.~Matthijs Jansen.
\newblock Fast spectroscopic imaging using extreme ultraviolet interferometry.
\newblock {\em Optics Express}, 32(16):28644--28654, 2024.

\bibitem{Wolf:82}
Emil Wolf.
\newblock New theory of partial coherence in the space--frequency domain. part
  i: spectra and cross spectra of steady-state sources.
\newblock {\em J. Opt. Soc. Am.}, 72(3):343--351, Mar 1982.

\bibitem{25}
Charles Bevis, Robert Karl, Jonathan Reichanadter, Dennis~F. Gardner, Christina
  Porter, Elisabeth Shanblatt, Michael Tanksalvala, Giulia~F. Mancini, Henry
  Kapteyn, Margaret Murnane, and Daniel Adams.
\newblock Multiple beam ptychography for large field-of-view, high throughput,
  quantitative phase contrast imaging.
\newblock {\em Ultramicroscopy}, 184:164--171, 2018.

\bibitem{29}
Andrew~M Maiden, Martin~J Humphry, and John M 
\newblock Ptychographic transmission microscopy in three dimensions using a
  multi-slice approach.
\newblock 29(8):1606--1614, 2012.

\bibitem{23}
Dennis~F. Gardner, Bosheng Zhang, Matthew~D. Seaberg, Leigh~S. Martin,
  Daniel~E. Adams, Farhad Salmassi, Eric Gullikson, Henry Kapteyn, and Margaret
  Murnane.
\newblock High numerical aperture reflection mode coherent diffraction
  microscopy using off-axis apertured illumination.
\newblock {\em Optics Express}, 20(17):19050--19059, 2012.

\bibitem{15}
Pengming Song, Ruihai Wang, Jiakai Zhu, Tianbo Wang, Zichao Bian, Zibang Zhang,
  Kazunori Hoshino, Michael Murphy, Shaowei Jiang, Chengfei Guo, and Guoan
  Zheng.
\newblock Super-resolved multispectral lensless microscopy via angle-tilted,
  wavelength-multiplexed ptychographic modulation.
\newblock {\em Optics Letters}, 45(13):3486--3489, 2020.

\bibitem{28}
Minh Pham, Arjun Rana, Jianwei Miao, and Stanley 
\newblock Semi-implicit relaxed douglas-rachford algorithm (sdr) for
  ptychography.
\newblock 27(22):31246--31260, 2019.

\bibitem{26}
Tsumoru Shintake.
\newblock {\em Phys. Rev. E}, 78:041906, Oct 2008.

\bibitem{27}
Chi-Feng Huang, Wei-Hau Chang, Ting-Kuo Lee, Yasumasa Joti, Yoshinori Nishino,
  Takashi Kimura, Akihiro Suzuki, Yoshitaka Bessho, Tsung-Tse Lee, and Mei-Chun
\newblock Xfel coherent diffraction imaging for weakly scattering particles
  using heterodyne interference.
\newblock 10(5), 2020.

\bibitem{30}
Hyeon-ho Son and Kyunghwan Oh.
\newblock Light propagation analysis using a translated plane angular spectrum
  method with the oblique plane wave incidence.
\newblock {\em Journal of the Optical Society of America A}, 32(5):949--954,
  2015.

\end{thebibliography}

\end{CJK}
\end{document}